\documentclass[preprint,aps,draft]{revtex4}
\usepackage{epsf}
\begin{document}
\title{Comment on "Erratum: Collective modes and gapped momentum states in liquid Ga: 
Experiment, theory, and simulation"
}

\author{Taras Bryk$^{1,2}$,
        Ihor Mryglod$^{1}$,
        Giancarlo Ruocco$^{3,4}$}

\affiliation{ $^1$ Institute for Condensed Matter Physics,National
Academy of Sciences of Ukraine,\\UA-79011 Lviv, Ukraine}
\affiliation{$^2$Institute of Applied Mathematics and Fundamental
Sciences,\\Lviv National Polytechnic University, UA-79013 Lviv,
Ukraine } 
\affiliation{$^3$
Center for Life Nano Science @Sapienza, Istituto Italiano di
  Tecnologia, 295 Viale Regina Elena, I-00161, Roma, Italy}
\affiliation{$^4$ Dipartimento di Fisica, Universita' di
Roma "La Sapienza", I-00185, Roma, Italy} 

\date{\today}
\begin{abstract}
We show, that the theoretical expression for the dispersion of collective excitations reported 
in  [Phys. Rev. B {\bf 103}, 099901 (2021)], at variance with what was claimed in the paper, does 
not account for the energy fluctuations and does not tend in the long-wavelegth limit to the 
correct hydrodynamic dispersion law.
\end{abstract}

\maketitle

Recently in \cite{Bry21} we reported on different mistakes in the theoretical scheme suggested 
in \cite{Khu20} for the calculation of the dispersion law of longitudinal collective excitations 
in liquids. Specifically, in their theoretical approach two types of fluctuations of conserved 
quantities, 
namely density and energy fluctuations, were missing that resulted in senseless expressions for the
current spectral function and dispersion of collective excitations. In response to our criticism 
the authors published an Erratum \cite{Khu21}  to their 
original paper explaining that "the equations referred to originally were carried over from an 
earlier version of the paper in error" \cite{Khu21}. We expected, therefore, that the correct 
theoretical expressions were published in the Erratum. On the contrary, in \cite{Khu21} the 
authors reported expressions identical as in an earlier paper \cite{Mok18} confirming in 
\cite{Khu21} that "the equations used were from our earlier theory". The theoretical method 
from \cite{Mok18} is a simple continued fraction representation of the
single Laplace-transformed density-density time correlation function ${\tilde F}_{nn}(k,z)$,  
which is equivalent to the standard improvement of the short-time behavior of the density-density 
time correlation function $F_{nn}(k,t)$ by an account for exact sum rules up to some order  
defined by the level of terminating the recursion in the continued fraction. The approach and 
closure relation for terminating the recursion in the continued fraction in \cite{Mok18} do not 
make explicit use of energy fluctuations.  
It is difficult to understand why the authors of  \cite{Khu20} and \cite{Khu21} claim 
in \cite{Khu21} that an account for energy fluctuations was made in their scheme.

To emphasize the absence of energy fluctuations in the expression reported in  \cite{Khu21} 
we make reference to the Generalized Collective Modes (GCM) theory for simple and many-component 
liquids, a generalised hydrodynamic
approach developed by some of us \cite{Mry95,Bry97,Bry01}. The approach is based on an extented set of dynamic variables (hydrodynamic plus orthogonal to them extended ones) \cite{deS88}. The extended set of dynamic variables (which includes energy density and its derivatives) was used for a matrix form of the generalized Langevin equation, solved in terms of dynamic  eigenmodes. The account for energy (or heat) fluctuations leads to simultaneous corresponding 
account for sum rules of the energy-density and energy-energy correlations with known long-wavelength
asymptotes. One can find out in \cite{Bry10} how the thermo-viscoelastic model and corresponding
analytical expressions can be reduced to the ones without account for energy (or heat) 
fluctuations.  In \cite{Khu21} the authors claimed that their expressions for the dynamic 
structure factor $S(k,\omega)$ (their Eq.12) and for the dispersion of collective excitations 
(their Eq.15), which were identical to those reported in \cite{Mok18} for a single continued fraction
representation of ${\tilde F}_{nn}(k,z)$, accounted for energy fluctuations. Let us perform some 
simple checks.

In \cite{Mok18} a set of dynamic variables was the following 
one: $A_0(k,t)=n(k,t)$, 
$A_1(k,t)=i{\bf k}\cdot {\bf J}(k,t)$, $A_2(k,t)=\dot{A}_1(k,t)+\Delta^2_1(k)A_0(k,t)$, ... , 
continuing this sequence by the next properly orthogonalized 
time derivatives of $A_2(k,t)$. In fact, this set of dynamic variables 
contained only two hydrodynamic ones of space-Fourier components of 
particle density $n(k,t)$ and of longitudinal component of current density $J^L(k,t)$, plus 
its time derivatives (extended variables).
The "relaxation parameters" 
$$\Delta^2_i(k)=\frac{\langle A_i(-k)A_i(k)\rangle}
{\langle A_{i-1}(-k)A_{i-1}(k)\rangle}$$ \cite{Mok18} 
have the known long-wavelength asymptotes: 
\begin{equation}\label{deltak0_1}
\Delta^2_1(k\to 0)\to c_T^2k^2,\qquad  \Delta^2_2(k\to 0)\to (c_{\infty}^2-c_T^2)k^2,\qquad 
\Delta^2_3(k\to 0)\to const~,
\end{equation}
and higher ones tending to corresponding constants in that limit. In (\ref{deltak0_1}) 
$c_T$ and $c_{\infty}$ are the isothermal and the high-frequency speeds of acoustic modes. 
If one applied these asymptotes to the expression for the dispersion of correlative excitations,
presented in \cite{Mok18} and \cite{Khu21}, it would result in 
$$
\omega(k\to 0)\to c_Tk~,
$$
with the isothermal speed of sound, but not being the correct hydrodynamic 
dispersion law $\omega(k\to 0)\to c_sk$ with the adiabatic speed of sound.
This is the standard drawback for the theories missing the coupling with 
energy  fuctuations. Consequently, the approach of \cite{Mok18} is not able to 
recover the correct origin (and the correct half-width) of the central 
Rayleigh peak of $S(k,\omega)$ in the long-wavelength region, which in the 
hydrodynamic regime for one-component liquids comes solely from thermal 
fluctuations.

However, in \cite{Khu21} the authors reported that a different set of dynamic
variables $W_n(k)$, containing the energy density, led to the same (!) expressions for 
the 
dynamic structure factor $S(k,\omega)$ and for the dispersion of collective 
excitations $\omega(k)$ in terms of "relaxation parameters" as in \cite{Mok18}. 
Remarkably, that in \cite{Khu21} the authors do not provide the explicit 
expressions for their set of 
dynamic variables, mentioning only that "... $W_n(k)$ are the dynamical 
variables associated with the particle density, longitudinal momentum 
component, energy density, and so on ...". Perhaps "... and so on ..." 
means the first and higher time derivatives of the $W_1(k)$ and $W_2(k)$ properly 
orthogonalized, however, this remains unclear being without any explanation in \cite{Khu21}.  
Their "relaxation parameters"
are expressed via $W_n(k)$ as follows
\begin{equation}\label{deltaW}
\Delta^2_n(k)=\frac{\langle W_n(-k)W_n(k)\rangle}
{\langle W_{n-1}(-k)W_{n-1}(k)\rangle}~\qquad n=1,2,3,4.
\end{equation}

Even before making any further checks one can say immediately, that from the point
of view of methodology of the continued fraction representation for the 
density-density time correlation functions\cite{Cop75} the $\Delta^2_2(k)$ must be 
related to the ratio of the fourth to the second frequency moments of the dynamic structure factor.
However, according to (\ref{deltaW}) and to the sequence of dynamic variables assured as
used in \cite{Khu21} the $\Delta^2_2(k)$ is instead directly related to the ratio of the 
energy-energy to 
the current-current correlators, that does not make any sense.
Nevertheless, let us first try to check the long-wavelength asymptote of the 
dispersion law 
(Eq.15) in \cite{Khu21} having the first two dynamic variables exactly the same
as in \cite{Mok18}, $W_0(k,t)=A_0(k,t)$, $W_1(k,t)=A_1(k,t)$, and the 
third one being the energy density $e(k,t)$ othogonalized to the density of particles
$$
h(k,t)=e(k,t)-n(k,t)\frac{\langle n(-k)e(k)\rangle}{\langle n(-k)n(k)\rangle}~,
$$
i.e. $W_2(k,t)=h(k,t)$. One can immediately notice, that there exists a problem 
with units of the "relaxation parameters", which in the standard continued 
fraction all must be of the same dimension of $[1/time^2]$. However, here the 
"relaxation parameter"
\begin{equation}\label{delta2_lim}
\Delta^2_2(k)=
\frac{\langle W_2(-k)W_2(k)\rangle}
{\langle W_{1}(-k)W_{1}(k)\rangle}\equiv
\frac{\langle h(-k)h(k)\rangle}
{k^2\langle J^L(-k)J^L(k)\rangle}\stackrel{k\to 0}{\to}\frac{mC_VT}{k^2}~, 
\end{equation}
i.e. has the dimension of $[energy^2*time^2]$. In (\ref{delta2_lim}) $C_V, T, m$ are 
specific heat at constant volume, temperature and atomic mass, respectively. 
The long-wavelength limit 
of $\Delta^2_2(k)$ diverges as $k^{-2}$, that immediately causes divergence of 
the ${\mathcal{A}}_2(k)$ (expressions after Eq.12 in \cite{Khu21}), which contains 
$\Delta^4_2(k)$, and other correlators cannot reduce the divergence of ${\mathcal{A}}_2(k)$ 
in the long-wavelength limit. All this leads to a senseless result for dispersion 
law in Eq.15 of \cite{Khu21}.  

Another possibility for the check is an expression of $W_1(k)$, not explicitly specified  
in \cite{Khu21}, in the form $W_1(k)=J^L(k,t)$, i.e. exactly the longitudinal component 
of the mass-current density (as mentioned in \cite{Khu21}) 
without the multiplier $k$ as used in \cite{Mok18} (and which we considered
above). With this choice the long-wavelength asymptotes of the "relaxation parameters" are:
\begin{equation}\label{deltak0_2}
\Delta^2_1(k\to 0)\propto const,\qquad  \Delta^2_2(k\to 0)\propto const,\qquad 
\Delta^2_3(k\to 0)\to k^2~,\qquad \Delta^2_4(k\to 0)\to const~,
\end{equation}
where we assumed that the fourth and the fifth dynamic variables are ${\dot J}^L(k)$ and 
${\dot h}(k)$, properly orthogonalized. However, there is enough to look at the first 
two asymptotes in (\ref{deltak0_2}) to see that the expression for the dispersion of 
collective excitations (Eq.15 in \cite{Khu21}) leads to senseless results too.

Hence, we have shown, that the expression for the dispersion of collective excitations
reported in the Erratum \cite{Khu21} is not consistent with the sequence of dynamic 
variables claimed as used ones in \cite{Khu21}, and which was an argument for an account 
for energy/thermal 
fluctuations in the proposed theoretical scheme. We can conclude that there are no proofs 
of the claimed account for energy/thermal fluctuations needed to recover the hydrodynamic 
dispersion law and the thermal origin of the central peak of dynamic structure factors in 
long-wavelength region. The theoretical scheme \cite{Mok18} does not account for thermal 
fluctuations in the sequence of dynamic variables too and results in the long-wavelength 
dispersion law with isothermal speed of sound. We suggest the authors of \cite{Khu21} to provide
the explicit expressions for the first five dynamic variables used in their scheme with an account 
for energy fluctuations as well as to show the explicit expressions for the long-wavelength 
asymptotes of the corresponding "relaxation parameters". Without these clarifying one can
assume that the authors provided untrue information about the account for energy fluctuations
by claiming \cite{Khu21}:
"This shows that the correlation function has the expected low-frequency limit and that the 
coupling between energy and density is included in our theory".

Another problem we would like to discuss here.
In \cite{Bry21} we made a check of the theoretical scheme only. However, in response of the 
claim in \cite{Khu21} that "the experimental and modelling parts of the paper are unaffected" we
checked out the simulation part of \cite{Khu20}. We would like to stress that the used  
in \cite{Khu20} solely NPT ensemble cannot be applied to correct calculations of $k$-dependent 
quantities. In fact, in NPT simulations the volume of simulation box fluctuates in order to keep 
the average pressure at the needed value. Consequently 
the vectors ${\bf k}=2\pi/L(n_x,n_y,n_z)$ with integer numbers $\pm n_x,\pm n_y,\pm n_z$ 
consistent with periodic boundary conditions for a chosen configuration with the boxlength $L$ 
will not satisfy the periodic boundary conditions for any next one with the boxlength $L'$. Hence, 
since in \cite{Khu21} only NPT simulations were reported, there exist only two choices: either
in \cite{Khu20} the authors reported untrue information about the simulation ensembles, or 
all the $k$-dependent quantities including dynamic structure  factors $S(k,\omega)$ and longitudinal
 (L) or transverse (T) current spectral functions $C^{L/T}(k,\omega)$ are absolutely wrong and 
inconsistent with periodic boundary conditions. 
 

\end{document}